\documentclass[conference]{IEEEtran}
\IEEEoverridecommandlockouts
\usepackage{amsmath,amssymb,amsfonts}
\usepackage{algorithmic}
\usepackage{graphicx}
\usepackage{textcomp}
\usepackage{xcolor}
\usepackage{hyperref}
\usepackage{siunitx}
\usepackage{balance}
\def\BibTeX{{\rm B\kern-.05em{\sc i\kern-.025em b}\kern-.08em
 T\kern-.1667em\lower.7ex\hbox{E}\kern-.125emX}}
 
\usepackage{listings}
\usepackage{fancyvrb}
\usepackage{float}
\usepackage{subcaption}
\usepackage{multirow}

\usepackage[noadjust]{cite}

\newif\iffinal
\iffinal
 \newcommand\supun[1]{}
 \newcommand\vibhatha[1]{}
 \newcommand\niranda[1]{}
 \newcommand\chathura[1]{}
 \newcommand\amila[1]{}
 \newcommand\er[1]{}
 \newcommand\inpr[1]{}
 \newcommand\deprc[1]{}
 \newcommand\rrw[1]{}
 \newcommand\rpt[1]{}
 \newcommand\rcmt[1]{}
\else
 \newcommand\supun[1]{{\color{red}[ Supun: #1 ]}}
 \newcommand{\vibhatha}[1]{{\textcolor{teal}{[ Vibhatha: #1 ]}}}
 \newcommand{\niranda}[1]{{\textcolor{orange}{[ Niranda: #1 ]}}}
 \newcommand\chathura[1]{{\color{green}[ Chathura: #1 ]}}
 \newcommand\amila[1]{{\color{purple}[ Amila: #1 ]}}
 
 \newcommand\er[1]{{\color{magenta}[ \textbf{[ERROR]}: #1 ]}}
 \newcommand\inpr[1]{{\color{teal}[ \textbf{[IN-PROGRESS]}: #1 ]}}
 \newcommand\deprc[1]{{\color{brown}[ \textbf{[DEPRECATED]}: #1 ]}}
 \newcommand\rrw[1]{{\color{cyan}[ \textbf{[READY-TO-REVIEW]}: #1 ]}}
 \newcommand\rpt[1]{{\color{olive}[ \textbf{[REPITITIVE-MATERIAL]}: #1 ]}}
 \newcommand\rcmt[1]{{\color{darkgray}[ \textbf{[COMMENT-REQUIRED]}: #1 ]}}
\fi

\begin{document}

%\title{TOMOGAN-ETI: TomoGAN Image Enhancing Engine on Edge TPU} % \zliu{better to have a more general title}
%\title{Portable Image Restoration Using Deep Learning} 
\title{Data Engineering for HPC with Python}

\author{
\IEEEauthorblockN{Vibhatha Abeykoon\IEEEauthorrefmark{1}\IEEEauthorrefmark{2}\textsuperscript{\textsection},
Niranda Perera\IEEEauthorrefmark{1}\textsuperscript{\textsection},
Chathura Widanage\IEEEauthorrefmark{1}\textsuperscript{\textsection}, 
Supun Kamburugamuve\IEEEauthorrefmark{2}\textsuperscript{\textsection} \\
Thejaka Amila Kanewala\IEEEauthorrefmark{3}\textsuperscript{\textsection},
Hasara Maithree\IEEEauthorrefmark{6},
Pulasthi Wickramasinghe\IEEEauthorrefmark{1}, \\
Ahmet Uyar\IEEEauthorrefmark{2},
and Geoffrey Fox\IEEEauthorrefmark{1}\IEEEauthorrefmark{2}
}

\IEEEauthorblockA{\IEEEauthorrefmark{1}Luddy School of Informatics, Computing and Engineering, IN 47408, USA\\
\{vlabeyko,dnperera,pswickra\}@iu.edu}
\IEEEauthorblockA{\IEEEauthorrefmark{2}Digital Science Center, Bloomington, IN 47408, USA\\
\{cdwidana, skamburu, auyar, gcf\}@iu.edu}
\IEEEauthorblockA{\IEEEauthorrefmark{3}Indiana University Alumni, IN 47408, USA\\
thejaka.amila@gmail.com}
\IEEEauthorblockA{\IEEEauthorrefmark{6}Department of Computer Science and Engineering, University of Moratuwa, Sri Lanka\\
hasaramaithree.15@cse.mrt.ac.lk}
}

\maketitle
\begingroup\renewcommand\thefootnote{\textsection}
\footnotetext{These authors contributed equally.}
\endgroup

\begin{abstract}
% Data engineering has become a vital part of scientific workloads. With the exponential growth of datasets, the efficient pre-processing, transformation and movement of data are considered necessities. In this paper, we describe the necessity of high performance computing (HPC) resources for productive and effective data engineering. Our contribution focuses on increasing usability with Python and efficient execution with a C++ back-end running on HPC environments. Unlike existing state-of-the-art data engineering tools written purely in Python, our solution adopts high performance compute kernels in C++, boasting an in-memory representation with Cython-based Python bindings. In the core system we use MPI for distributed memory computations with a data parallel approach for processing large datasets in HPC clusters. For the experiments, we discuss the application of our high performance Python library compared to state-of-the-art data engineering solutions written in Python.

Data engineering is becoming an increasingly important part of scientific discoveries with the adoption of deep learning and machine learning. Data engineering deals with a variety of data formats, storage, data extraction, transformation, and data movements. One goal of data engineering is to transform data from original data to vector/matrix/tensor formats accepted by deep learning and machine learning applications. There are many structures such as tables, graphs, and trees to represent data in these data engineering phases. Among them, tables are a versatile and commonly used format to load and process data. In this paper, we present a distributed Python API based on table abstraction for representing and processing data. Unlike existing state-of-the-art data engineering tools written purely in Python, our solution adopts high performance compute kernels in C++, with an in-memory table representation with Cython-based Python bindings. In the core system, we use MPI for distributed memory computations with a data-parallel approach for processing large datasets in HPC clusters. 
\end{abstract}


\section{Introduction}

In the last two decades data has played a major role in the evolution of scientific discoveries. This impacts a wide range of domains such as medicine, security, nature, climate, astronomy, and physics. Massive amounts of data are collected daily by a plethora of scientific applications, especially with the evolution of the Internet of Things (IoT)\cite{castillo2015projecting,whitmore2015internet}. The transformation of raw data to a form suitable for analytics is a complex process known as \emph{data engineering} \cite{gray2000rules}. Data engineering involves many data formats, as well as the transformation, movement, and storage of said data. To support these operations a major contribution has been done by the big data community. Among these contributions, Apache Spark\cite{apache-spark}, Apache Flink\cite{apache_flink}, Apache Beam\cite{akidau2015dataflow}, Twister2\cite{twister2} and Hadoop\cite{apache-hadoop} can be considered as widely used systems for data engineering. 

Data scientists who design analytical models often use programming languages such as R, Matlab and Python. The flexibility of these languages offers ideal prototyping required for experiments. Additionally, the emergence of machine learning (ML) and deep learning (DL) frameworks such as Scikit-Learn\cite{pedregosa2011scikit}, PyTorch\cite{pytorch}, Tensorflow\cite{tensorflow} and MxNet\cite{mxnet} has inclined the data analytics domain to rely on Python. Although the core system of most prominent tools from the big data community are written in JVM-based languages (like Java and Scala), Python APIs are an essential feature for bridging the gap between data engineering and data analytic frameworks. PySpark, PyHadoop, PyFlink and PyTwister2 are some notable examples. This interfacing affects the efficiency of the system, since the data has to be serialized/deserialized back-and-forth the Python runtime and JVM runtime. There has been some efforts taken to improve this performance bottleneck by using columnar data formats like Apache Arrow\cite{apache-arrow}. But we observe that the performance can be further improved by using high performance compute kernels. 

Analytical engines increasingly rely on high performance computing clusters for training large deep learning models\cite{resnet50, naumov2019deep, devlin2018bert}. The data engineering frameworks must be able to leverage the full power of these clusters to feed the data efficiently to such applications. However, current big data systems are not directly compatible with HPC frameworks such as MPI\cite{open_mpi,MPI-3.0_2012}, PGAS\cite{zheng2014upc++} or Slurm. There are several data engineering frameworks contributed from the Python community. Python support is required to blend with most of the data analytical engines written with Python APIs. Pandas\cite{pandas}, Modin\cite{modin}, Dask\cite{dask} and Mars\cite{marsproj55:online} are a few examples of such well-known solutions written in Python. These frameworks are not enhanced with HPC-oriented compute kernels or with HPC-oriented distributed computing frameworks. This shows that there is an opportunity to design high performance data engineering libraries optimized for HPC resources. Further language bindings like Cython\cite{behnel2011cython} can be used to develop Python API around high performance kernels. 

\begin{figure*}[htpb]
\begin{center}
\includegraphics[width=0.85\textwidth]{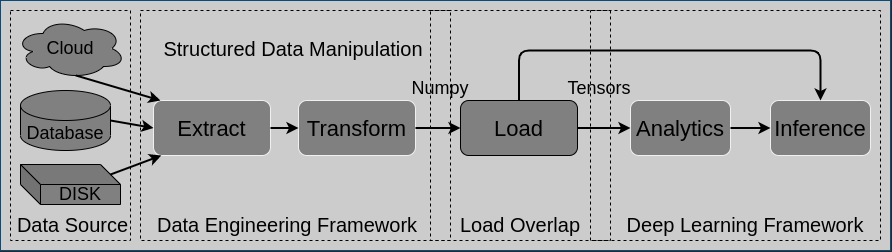}
\end{center}
\caption{Data Engineering to Data Analytics}
\label{fig:data-engineering-and-data-analytics}
\end{figure*}

Extending from the best practices of the big data and Python communities, we believe that data engineering can be enhanced to suit the performance demands of HPC environments. Since our effort is to seamlessly bridge data engineering and data analytics, we must understand the expectations of state-of-the-art data analytics engines. Among many data structures available for representing data in data engineering phases, tables are one of the most widely used abstractions. Relational algebraic operations are a natural fit for processing table data, and SQL interfaces can further enhance usability. Tables can be partitioned into distributed nodes, and the operations can work the partitioned tables. With an efficient data representation and with high performance compute kernels, we believe that data engineering and data analytics can be bridged efficiently and effectively. \autoref{fig:data-engineering-and-data-analytics} depicts the connection between the data engineering and the data analytic in a data pipeline.

In this paper we present a high performance Python API with a C++ core to represent data as a table and provide distributed data operations. We compare our work with existing data engineering libraries in Python and big data. The rest of the paper is organized as follows. Section \ref{s:data-engineering-and-hpc} discusses the data engineering paradigm with high performance computing. In Section \ref{s:cylon} we elaborate on architecture and implementation details, while in Section \ref{s:pycylon} demonstrates how Python-bindings are written on Cylon high performance compute kernels. In Section \ref{s:experiments} demonstrates how our implementation compares with existing data engineering solutions. Section \ref{s:related-work} reviews the existing literature. We reach our conclusions in Section \ref{s:conclusion} and \ref{s:future-work} with an analysis of the current progress and future goals in our effort to improve data engineering with Python in HPC environments.

% Modin, write about the Pandas optimization done by modin and mention why it still can have performance bottlenecks because of not having high performance compute kernels. 

% \vibhatha{Write more on lazy execution on Spark and Dask and elaborate the necessity of eager execution as proven by frameworks like PyTorch}

% \vibhatha{Points to add to intro}
% In modern science, data engineering has become a vital component on the road to design scientific solutions for 

% Discuss the relation to big data systems and data engineering (Apache Spark with PySpark), 
% Apache Flink, Twister2. 

% Disjoint nature between these big data systems and Deep Learning frameworks. Qualitative and Quantitative facts and existing benchmarks (ser-des cost, memory representation, Language limitations, etc).

% Discuss on Data Engineering best practices with the state of the art frameworks. 
% Functions and current limitations of these frameworks. (Pandas, Modin, Mars) 

% Discuss on high performance C++ libraries written to support data pre-processing. 
% (Apache Arrow) 

% \supun{Comment 1}\\
% \amila{Comment 2}\\
% \niranda{Comment 3}\\
% \chathura{Comment 4}\\
% \vibhatha{Comment 5}\\

\section{Data Engineering}\label{s:data-engineering-and-hpc}

The data used by ML/DL applications comes in the form of vectors, matrices or tensors. In most cases, the original data are not in formats that are directly convertible to the data structure expected by ML/DL applications. This data can be in many forms inside relational/graph/NoSQL databases or files. The path from raw to readable data is not a one-step process and can be highly complicated depending on the use case. Vector/matrix/tensor data structures are represented by arrays (contiguous memory) with homogeneous data values in the main memory.  

In contrast, depending on the original data format, there are many structures available to model them, such as tables, graphs and trees. It is fair to say that table is the most common structure used to represent data. A table contains a set of rows and columns viewed as a grid with cells. The columns can contain data of different types compared to a matrix. Natural operations for tables come from relational algebra. The fundamental relational algebra operations are shown in Table~\ref{tb:ops}.

\begin{table*}[t]
\centering
{\renewcommand{\arraystretch}{1.3} 
\begin{tabular}{|l|p{14cm}|}
\hline
Operator   & Description                                                                \\ \hline
Select     & Select operator works on a single table to produce another table by selecting a set of attributes matching a predicate function that works on individual records.                                                 \\ \hline
Project    & Project operator works on a single table to produce another table by selecting a subset of columns of the original table.                    \\ \hline
Join       & Join operator takes two tables and a set of join columns as input to produce another table. The join columns should be identical in both tables. There are four types of joins with different semantics: inner join, left join, right join and full outer join. \\ \hline
Union      & Union operator works on two tables with an equal number of columns and identical types to produce another table. The result will be a combination of the input tables with duplicate records removed.                                 \\ \hline
Intersect  & Intersect operator works on two tables with an equal number of columns and identical types to produce another table that holds only the similar rows from the source tables.                                       \\ \hline
Difference & Difference operator works on two tables with an equal number of columns and identical types to produce another table that holds only the dissimilar rows from both tables.                                            \\ \hline
\end{tabular}
}
\caption{Fundamental relational algebra operations\label{tb:ops}}
\end{table*}

To support large-scale data engineering on top of a table structure, we need to have an in-memory table representation, a partitioning scheme to distribute the table across nodes, and distributed data operations on the partitioned table. As with matrices, tables can be represented using a column or row format. They can also be partitioned row-wise, column-wise or using a hybrid approach with both column and row partitioning. The operations shown in Table~\ref{tb:ops} can be implemented on a distributed table partitioned into multiple compute nodes. Both big data and Python systems have used the table abstraction to process data. Another advantage in a table abstraction is that it can be queried using SQL. 

\subsection{Big Data Systems}

Big data systems adopt the dataflow model, where functional programming is heavily implemented to support a series of data operations. Figure \autoref{fig:pyspark-dataflow-operations} shows an example dataflow operation in PySpark. The big data systems are designed to run on commodity cloud environments. Referring to recent advancements in DL frameworks, eager execution (inverse of lazy execution) has been the most widely adopted programming model as it mimics the Python programming style. Eager execution is not supported by the Python bindings of existing big data systems such as PySpark, PyFlink, PyHadoop and Beam-Python (Apache Beam Python API), which adopt the dataflow model. Integrating dataflow frameworks with high performance computing resources is a challenging task. Frameworks such as Twister2\cite{twister2} and APIs like TSet\cite{wickramasinghe2019twister2} were developed to bridge this gap between the big data and HPC systems. Although these efforts enabled big data systems to run on HPC environments, working with multi-language runtimes produces a bottleneck. In such integrations, we observe the following common approaches in big data systems. 

\begin{itemize}
    \item JVM-based back-end handling computation and communication
    \item Python-based API for defining the dataflow application
\end{itemize}

\lstset{
   basicstyle=\fontsize{7}{7}\selectfont\ttfamily
}
\begin{figure}[htpb]
\caption{PySpark Dataflow Operations}
\begin{lstlisting}[language=Python]
df_r = sqlContext.read.format('com.databricks.spark.csv') \
        .options(header='true', inferschema='false') \
        .load(..).toDF(...).repartition(...).cache()
\end{lstlisting}
    \label{fig:pyspark-dataflow-operations}
\end{figure}

% \inpr{} \vibhatha{I think we can remove this para! -niranda}
Although JVM-Python bridging is done to enable switching between language runtimes, it involves data serialization and deserialization. This creates a performance bottleneck in large-scale applications since (a) it consumes a significant amount of additional CPU cycles for data serialization/deserialization, and (b) this approach also tends to create additional copies of data on both language runtimes' memory spaces, reducing the effective memory available for the application. With this setting, enabling high performance kernels for legacy big data frameworks poses a challenge.

\subsection{Python for Data Engineering}

PyData and Python community-designed frameworks emerged well ahead of the \emph{big data explosion}. Pandas was initially released in 2008, compared to Apache Spark in 2014. Its user-friendly front-end heavily influenced the front-end APIs of big data frameworks that followed. In the last decade, Pandas became the center for all data engineering tasks associated with ML/DL problems. It is predominantly written in Python and seamlessly integrates with ML/DL frameworks with Python front-ends. 

% Pandas, a fully-fledged Python solution for data engineering was introduced in 2008. This was a couple of years earlier to the dawn of Big Data systems (Apache Spark, Apache Flink, Hadoop, Apache Beam, etc). 
Pandas provided a rich API for data engineering and wraparound tabular format data processing for heterogeneous data, establishing a convenient way to preprocess the data. 
% Here the heterogeneous data refers to data with multiple types. The raw data extracted from the original data source doesn’t have the numerically pre-processed matrix-format data. 
Yet Pandas suffers from performance bottlenecks for several reasons. Firstly, it only works on a single core. This gives much less room to scale out and process larger datasets in parallel. Secondly, the compute kernels are entirely written in Python without high performance compute kernels. For higher performance and effective prototyping, Apache Spark introduced PySpark which includes a similar functionality. But it is also constrained by the data movement between the Python runtime and JVM. 
% For data scientists getting familiar with big data tools and system, setups are also challenging. 

Since Pandas first premiered, there have been some attempts made by the Python community to improve the performance for large-scale problems. A distributed dataframe \cite{lazardask} was introduced by Dask. This is a full-fledged Python library for parallel computing and it allows for scaling the dataframe to a larger number of machines. The API in Dask is based on dataflow-like lazy execution that uses Python Futures.
% Here a task graph is being internally generated depending on the data operations and they are executed once the execute method is called upon a given operator. This will execute all the prior operations connected to this operator in order. Also, Dask supports Python Futures as responses and results can be obtained once the tasks are completed asynchronously or block if results are requested. Dask supports distributed computing using a client-server architecture. Also, Dask doesn’t support high performance compute kernels. 

Extending from Pandas dataframes, Modin\cite{modin} optimizes the query compiling and internal components of Pandas. It also supplies distributed computation by using Dask or Ray as the execution back-end. The internal optimizations have provided better performance over Pandas. Similar to Modin, a framework called Mars \cite{marsproj55:online} was released by Alibaba with a Ray-based back-end. Both these frameworks are developed in Python and do not support high performance compute kernels.

\subsection{HPC for Data Engineering}

% Although these frameworks are compatible and seamlessly integrates with the data analytic engines, the usage of high performance kernels for internal data operations are not supported yet. This necessity transformed into a vital requirement with the increasing data and computation intensive deep learning applications. This brings in the requirement to evolve and shift the existing data engineering paradigm towards HPC. HPC resources provide multiple advantages for data engineering. When processing large datasets, distributed memory computations are vital to get better performance by scaling in HPC clusters. 

We observe that ML/DL applications commonly consume large predefined and preprocessed datasets. They frequently use data-parallel execution mode in distributed execution. This indicates that data engineering and ML/DL execution occur as two separate functions. We believe that there is an opportunity to bridge this gap using a high-performance low-overhead framework for data engineering, thereby increasing the efficiency and performance of the ML/DL pipeline. We believe introducing HPC for data engineering would enable this transformation. Data engineering in GPU resources using Cudf \cite{cudf} dataframes is a good example of such an effort, as it delivers processed data immediately to the ML/DL pipeline.

Data engineering would benefit by using distributed memory computations. Furthermore, there are well-defined HPC compute kernels for numerical data analysis. BLAS\cite{blackford2002updated}, MKL\cite{wang2014intel} and Boost\cite{schaling2011boost} are a few examples of such tools being regularly used. Developing systems with these HPC resources provides a better opportunity to improve the performance in data engineering. In addition, HPC compute kernels written in Fortran/C/C++ can be easily integrated with Python, allowing for a seamless integration among HPC resource, data engineering and data analytics. 

%  And most of these libraries are not optimized to run in the state of the art HPC resources. In HPC systems, most of the prominent scientific applications are written with MPI and PGAS. Also, large scale distributed job management is also done with resource schedulers like Slurm, Aprun and Torque. \vibhatha{write something about HPC specific storage mechanisms}. Developing HPC solutions for data engineering requires to be associated with these resources. Since the data engineering is a field mostly associated with the Big Data and Python community, most of the existing work is not fully-fledged HPC-based Python solutions. We realize this as the main aspect where performance can be boosted to provide high performance data engineering with Python. 
Since high performance Python is tightly coupled with kernel implementations written in C/C++, the compute kernels related to data engineering can also be written in C/C++. Python bindings must be written carefully without degrading performance and usability. Frameworks like Swig \cite{beazley1996swig}, Pybind11 \cite{jakob2017pybind11} and Cython \cite{behnel2011cython} are mainly used to write efficient Python bindings. In building frameworks or complex applications, the most recommended methods are Python bindings written in either Pybind11 or Cython. This is evident from open source tools such as PyTorch(PyBind11), Numpy(Cython) and Cudf(Cython). Pybind11 favors more on programming with C++ 11 standards, while Cython focuses on both C++ and Python approaches. 

\begin{figure}[htb]
\begin{center}
\includegraphics[width=0.35\textwidth]{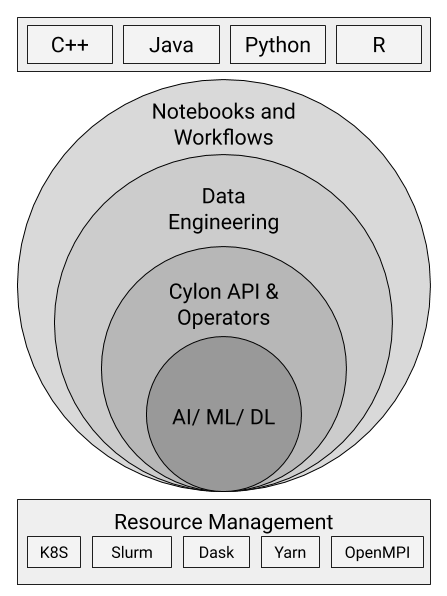}
\end{center}
\caption{PyCylon's position in data engineering}
\label{fig:cylon-position-data-engineering}
\end{figure}

The aforementioned tools are mainstream CPU data engineering engines. Rapids AI developed CuDF \cite{cudf}, which offers a GPU dataframe written on top of high performance GPU kernels. CuDF houses a Cython-based Python binding on top of the core GPU dataframe API. Similar to other Python-based frameworks, CuDF uses Dask as a back-end for distributed execution. Limited memory availability in GPUs is often seen as the main drawback for CuDF. Even though GPUs offer much faster kernel computations, CPU-based solutions are still the best fit for big data-related data engineering. Furthermore, we observe that the CPU manufacturers are constantly improving CPU hardware architectures with more threads, faster cache memory, and low power consumption. Therefore a CPU-based solution may provide a less expensive alternative for larger dataset operations. 

\subsection{Jupyter Notebooks}\label{s:s:jupyter-integration}

\begin{figure}[htb]
\begin{center}
\includegraphics[width=0.35\textwidth]{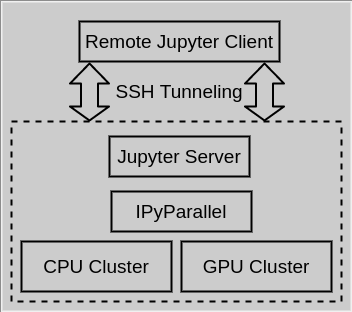}
\end{center}
\caption{Cylon Jupyter Cluster}
\label{fig:cylon-jupyter-cluster}
\end{figure}

Jupyter Notebooks is a powerful means to share and maintain data engineering workflows. When it comes to distributed workloads, Jupyter Notebooks does not support running programs in a cluster with the interactive look and feel generally provided for sequential programs. For Cylon we use an existing set of plugins created by the Python community to enable this on Jupyter Notebooks, with IPyParallel\cite{ipythoni44:online} and a distributed extension to IPython\cite{perez2007ipython} kernels. This distributed integration currently supports only an MPI cluster. Using this integration, a data engineering job can be seamlessly integrated with distributed ML/DL jobs. \autoref{fig:cylon-jupyter-cluster} shows the architecture of the current implementation.  

\section{Cylon}\label{s:cylon}

Having recognized the cost of performance bottlenecks when designing a high performance data engineering library, we created a library/framework Cylon\footnote{https://github.com/cylondata/cylon}. Cylon a is a distributed memory data table consisting of core relational algebra operators written in C++ high performance kernels. \autoref{fig:cylon-position-data-engineering} shows PyCylon's position on data engineering. 

\subsection{Data Model}\label{s:s:data-model}
% \vibhatha{Does this need re-writing... very much similar to last paper. I just added it here and did a minor modofication.}
Generally data processing systems are divided into two categories: (1) Online Transaction Processing (OLTP) and (2) Online Analytical Processing (OLAP). OLTP usually deals with a large number of atomic operations such as inserts, updates, and deletes, while OLAP is focused on bulk data processing (bulk reads and writes) with a low volume of transactions. Cylon falls into the OLAP category. For the in-memory data representation, Cylon follows the Apache Arrow format. Apache Arrow uses columnar in-memory representation. This allows Cylon to seamlessly integrate with Arrow-based frameworks like Pandas, Apache Spark, Modin, Parquet and Numpy. 

\subsection{Distributed Memory Execution}\label{s:s:distributed-memory-execution}

In order to handle massive volumes of data which do not fit into the memory of a single machine, Cylon employs distributed memory execution techniques to slice a large table into small pieces across a cluster of computing nodes. Applying an operation on a table applies that operation concurrently across all the table partitions that reside on different nodes. Hence Cylon embraces the data-parallel approach to parallelizing the computing tasks. Additionally, the Cylon worker process spawns only a single thread for execution. We assume that the user will start multiple instances of Cylon workers within a single node to match the number of available CPU cores and thus fully utilize the computing capacity of the hardware.

\subsection{Operators}\label{s:s:operators}

Cylon provides a set of operators for communication-related to distributed computing and relational algebra operators for processing tabular data. The set of operations that we have implemented in Cylon requires an additional communication step when running in a distributed environment. Cylon performs a key-based partition followed by a key-based shuffle through the network to collect similar records into a single process. We have implemented an AllToAll communication utilizing the asynchronous send and receive capabilities of the underlying communication framework to support this cause. The initial implementation of Cylon is written with OpenMPI to handle the communication, which can be easily pluggable with a different framework such as UCX. Cylon can utilize the RDMA capabilities or any other hardware-level network accelerators supported by OpenMPI to improve CPU utilization, throughput, and the latency of the distributed operations. When running on OpenMPI, apart from calling Cylon's built-in functions to manipulate data on the tables, users are free to do any OpenMPI API call at any point of execution to handle additional computing or message passing requirements.
Also, Cylon supports a set of core relational algebra operators used in manipulating tabular data. These are supported as local and distributed operators. The core data table operators are shown in Table \ref{tb:ops}.

\section{PyCylon}\label{s:pycylon}

\begin{figure}[htb]
\begin{center}
\includegraphics[width=0.45\textwidth]{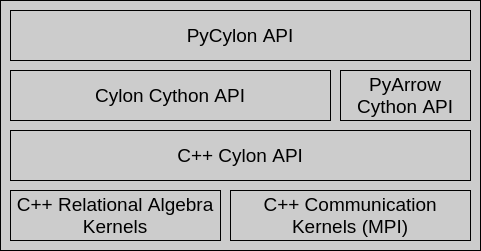}
\end{center}
\caption{Cylon API Overlay}
\label{fig:cylon-api-overlay}
\end{figure}

PyCylon is the Python API written on top of Cylon high performance kernels. Cython is used to link the C++ kernels with Python. The core compute kernels for relational algebra operations are written in C++ using the same memory buffer allocated in C++ when doing computations from the Python API. Since the in-memory data representation is based on Apache Arrow, we have also extended our support to PyArrow tables via our Cython API. This seamlessly integrates the computations from PyArrow extended libraries to PyCylon. We do not expose the communication API to the data scientist; instead the communications are internally handled when doing distributed computations on data tables. \autoref{fig:cylon-api-overlay} shows the API overlay of Cylon. 

Even though Cylon is a distributed memory framework, it can also be used as a high performance library to speed up the data processing workloads written in similar Python libraries like Pandas, Modin and Dask. PyCylon includes a DataTable API which is being continuously improved to provide advanced functionality to the data engineering workloads. 

\begin{figure}[htbp]
\begin{center}
\includegraphics[width=0.45\textwidth]{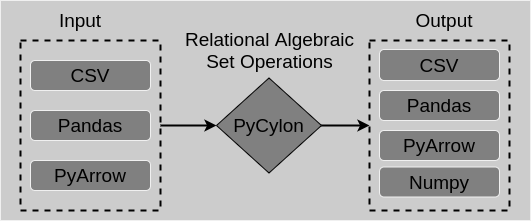}
\end{center}
\caption{Cylon Data Interoperability}
\label{fig:cylon-data-interoperability}
\end{figure}

\autoref{fig:cylon-data-interoperability} shows the PyCylon data interoperability. PyCylon currently supports a few input data formats and input data libraries. CSV support is provided via core Cylon C++ and with PyArrow and Pandas. A PyCylon table can also be created by passing a Pandas dataframe or a PyArrow table. Once the distributed computation is complete, the output can be converted to CSV, Pandas dataframe, etc. 

\subsection{PyCylon Table}\label{s:s:pycylon-table}

The higher level API we have included is the DataTable API. This provides parallelism-unaware API endpoints so that a data scientist can prototype the model without worrying about complex parallel computing concepts. The Cylon context manages the initialization of a distributed environment, so the data scientist need only specify the back-end name used (i.e MPI or UCX). Currently our implementation supports MPI.

\lstset{
   basicstyle=\fontsize{7}{7}\selectfont\ttfamily
}
\begin{figure}
\caption{PyCylon Sequential Join}
\begin{lstlisting}[language=Python]
from pycylon.data.table import csv_reader
from pycylon.data.table import Table
from pycylon.ctx.context import CylonContext

ctx: CylonContext = CylonContext()

tb1_file = f'{/path/to/file_1}'
tb2_file = f'{/path/to/file_2}'

tb1: Table = csv_reader.read(ctx, tb1_file, ',')
tb2: Table = csv_reader.read(ctx, tb2_file, ',')

configs = {'join_type':'left', 'algorithm':'hash', 
                'left_col':0, 'right_col':0}
                
tb3: Table = tb1.join(ctx, table=tb2, 
        join_type=configs['join_type'], 
        algorithm=configs['algorithm'],
        left_col=configs['left_col'], 
        right_col=configs['right_col'])
        
tb3.show()
ctx.finalize()
\end{lstlisting}
    \label{fig:pycylon-join-sequential}
\end{figure}

\lstset{
   basicstyle=\fontsize{7}{7}\selectfont\ttfamily
}
\begin{figure}
\caption{PyCylon Distributed Join}
\begin{lstlisting}[language=Python]
from pycylon.data.table import csv_reader
from pycylon.data.table import Table
from pycylon.ctx.context import CylonContext

ctx: CylonContext = CylonContext("mpi")

tb1_file = f'file_1_{ctx.get_rank()}'
tb2_file = f'file_2_{ctx.get_rank()}'

tb1: Table = csv_reader.read(ctx, tb1_file, ',')
tb2: Table = csv_reader.read(ctx, tb2_file, ',')

configs = {'join_type':'left', 'algorithm':'hash', 
                'left_col':0, 'right_col':0}
                
tb3: Table = tb1.distributed_join(ctx, table=tb2, 
        join_type=configs['join_type'], 
        algorithm=configs['algorithm'],
        left_col=configs['left_col'], 
        right_col=configs['right_col'])
        
tb3.show()
ctx.finalize()

\end{lstlisting}
    \label{fig:pycylon-join-distributed}
\end{figure}

The difference between a sequential and distributed table API call is the distributed prefix for the method name, ‘mpi’ argument for the context initialization, and specifying unique file names for each process. In the current API we have kept the MPI look and feel so that users can extend PyCylon programs to existing MPI projects or vice versa. We also support table conversions and initialization as follows. Figures \ref{fig:pycylon-join-sequential} and \ref{fig:pycylon-join-distributed} show the PyCylon code for sequential and distributed join respectively. Figure \ref{fig:pycylon-conversions} displays the code-wise data interoperability previously illustrated in Figure \ref{fig:cylon-data-interoperability}.

\lstset{
   basicstyle=\fontsize{7}{7}\selectfont\ttfamily
}
\begin{figure}
\caption{PyCylon Conversions}
\begin{lstlisting}[language=Python]
# pycylon_table = Table.from_arrow(pyarrow_table),
# pycylon_table = csv_reader.read(...) 
tb3: Table = <pycylon_table> 
# converts PyCylon table to Pandas
pdf: pd.DataFrame = tb3.to_pandas()
# converts PyCylon table to Numpy, specify 
npy: np.ndarray = tb3.to_numpy(order='C')
\end{lstlisting}
    \label{fig:pycylon-conversions}
\end{figure}

We continue to improve the API endpoints and are adding more functionality for users. We currently offer an experimental Numpy support with PyCylon. The Numpy conversion is the direct link to the tensors in deep learning. In PyTorch Numpy especially, NdArray to tensor conversion takes a negligible amount of time. 

\section{Experiments}\label{s:experiments}

% Distributed Join and Union experiments with C++/Python PyCylon. 
% Discuss the performance and overheads. 
% Add Modin vs PyCylon distributed join,union. 
% Add Pandas vs PyCylon single core join,union.

We analyzed the strong scaling performance of PyCylon for the following scenarios and compared the performance against popular Python-based frameworks Dask (Distributed), Modin, and PySpark. 

\begin{enumerate}
    \item Strong scaling performance comparison between the frameworks. Join operation was used here as it is a common use case of columnar-based traversal.
    \item Larger test with the best performing frameworks of the above experiment.
    \item Overhead comparison between Cylon's Python and Java bindings. 
\end{enumerate}

\emph{Hardware Setup}: The tests were carried out in a cluster with 10 nodes. Each node is equipped with Intel\textsuperscript{\textregistered} Xeon\textsuperscript{\textregistered} Platinum 8160 processors. A node has a total RAM of 255GB and mounted SSDs were used for data loading. Nodes are connected via Infiniband with 40Gbps bandwidth. 

\emph{Software Setup}: Cylon was built using g++ (GCC) 8.2.0 with OpenMPI 4.0.3 as the distributed runtime. \emph{Mpirun} was mapped by nodes and bound sockets. Infiniband was enabled for MPI. For each experiment, a maximum of 40 cores from each node were used, reaching a maximum parallelism of 400.

For the baseline sequential experiments we used Pandas 0.25.3 version. Apache Spark 2.4.6 (hadoop2.7) pre-built binary was chosen for this experiment alongside its PySpark release. Apache Hadoop/ HDFS  2.10.0 acted as the distributed file system for Spark, with all data nodes mounted on SSDs. Both Hadoop and Spark clusters shared the same 10-node cluster. To match MPI setup, \emph{SPARK\_WORKER\_CORES} was set to 16 and \emph{spark.executor.cores} was set to 1. Additionally we also tested PySpark with \emph{spark.sql.execution.arrow.pyspark.enabled} option, which would allow PyArrow underneath PySpark dataframes. 

Dask and Dask-Distributed 2.19.0 was set up with Pip installation. Dask Distributed cluster remained in the same nodes as mentioned previously, with \emph{nthreads=1} and varying \emph{nprocs} based on the parallelism. All workers were equally distributed among the nodes.

Modin \cite{modin} 0.6.3 was selected alongside Ray 0.7.3 for the experiments. Note that both these frameworks are several versions behind the latest released versions. We were unable to get Modin's latest version (0.8.0) to work with its corresponding Ray 0.8.7 back-end. At the time of conducting these experiments, there are a number of Github issues reported on the same incident which have not yet been resolved \cite{ray-issues1}\cite{ray-issues2}. 

\emph{Dataset Formats}: For strong scaling test cases, CSV files were generated with four columns (one int\_64 as index and three doubles). The same files were then uploaded to HDFS for the Spark setup and output counts were checked against each other to verify the accuracy. Timings were recorded only for the corresponding operation (no data loading time considered). For the extended test case, CSV files with two columns were used  (one int\_64 as index and one double as payload).

\subsubsection{Scalability}

\begin{figure}[htb]
\begin{center}
\includegraphics[width=0.50\textwidth]{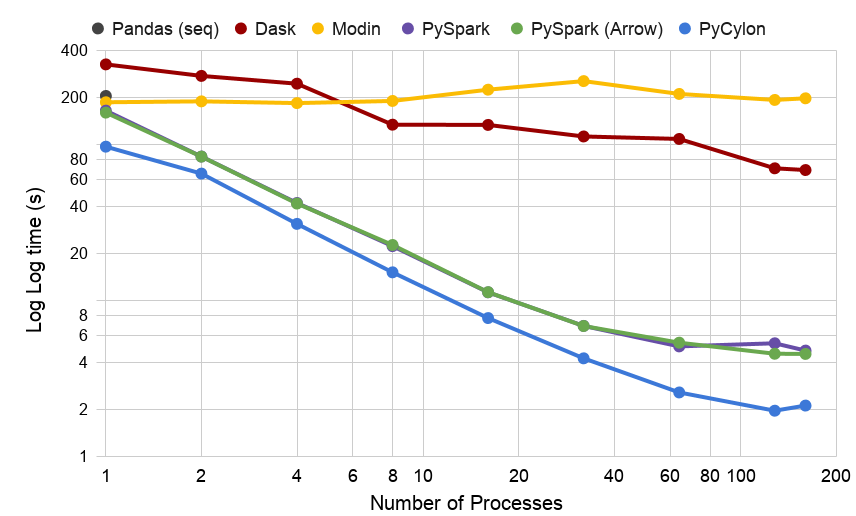}
\end{center}
\caption{Strong Scaling with Join Operator. Distributed Join operation was called across 1-160 processes across 10 Nodes (Log-Log plot)}
\label{fig:strong-scaling-join-operator}
\end{figure}

To test the scalability of the Python data engineering frameworks, we varied the parallelism from 1 to 160 while keeping total work at 200 million rows per relation (left/ right). The results are shown in Figure \ref{fig:strong-scaling-join-operator}. 

According to our findings, PyCylon seemed to scale well as the number of processes increased. Around 160 processes, the speedup reaches its plateau. This was expected, as when the parallelism increases, the operation transforms into a communication-bound operation. These results coincide with the Cylon C++ performance in our prior publication \cite{widanage2020high}.

Additionally, the current Cylon compute kernels do not take into account factors such as NUMA boundaries, in-cache performance, etc. As the number of processes inside a node increases, we can expect resource contention for memory bandwidth and L1/L2 caches. Polychroniou et al \cite{polychroniou2014comprehensive} show that these factors play a vital role in sorting operations, which is the core task in Cylon joins. 

Out of Dask-distributed, Modin, and PySpark, only PySpark achieved the strong scaling we expected. This reaffirms the adoption of Spark as a popular data engineering tool. Even with PyArrow execution enabled, the performance seemed to be identical.  

Dask-distributed shows some strong scaling conformity, but since it is developed with a Python back-end, this behavior is nothing out of the ordinary. 

Recently Modin \cite{modin} emerged as a distributed dataframe successor for Pandas. Authors Pertersohn et al showed that it was able to work on datasets exceeding 100GB \cite{petersohn2020towards}. We have tested Modin's experimental distributed execution with Ray back-end, but found it performs poorly for strong scaling. Even though Modin's (nearly) complete Pandas dataframe API looks very promising, it suggests that there is a lot of room for improvement.

\subsubsection{Larger Load Tests}
\begin{figure}[htb]
\begin{center}
\includegraphics[width=0.50\textwidth,height=0.28\textwidth]{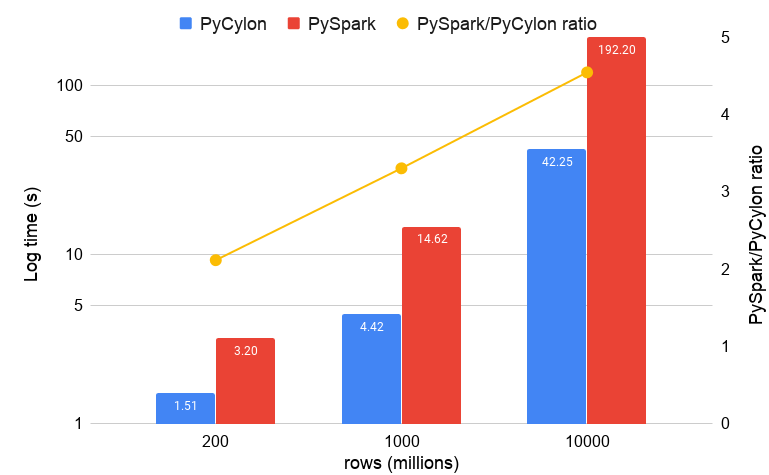}
\end{center}
\caption{PyCylon vs. PySpark for Joins with 200 processes at each experiment (Log scale on vertical axis and labels on horizontal axis)}
\label{fig:cylon-spark}
\end{figure}

From the above experiment, PySpark and PyCylon scaled as expected. As such we carried out a larger test to determine how these two frameworks perform under bigger loads. We fixed the processes at 200 and varied the total work from 200 million rows per relation to 10 billion. The results are shown in Figure \ref{fig:cylon-spark}. 

As the total work increases from 200 million rows to 10 billion, the ratio between PySpark time and PyCylon time increases from 2.1 to 4.5 times. This indicates that Cylon performs better at larger workloads. 

\subsubsection{Switching Between C++, Python, and Java}
\begin{figure}[htb]
\begin{center}
\includegraphics[width=0.50\textwidth,height=0.28\textwidth]{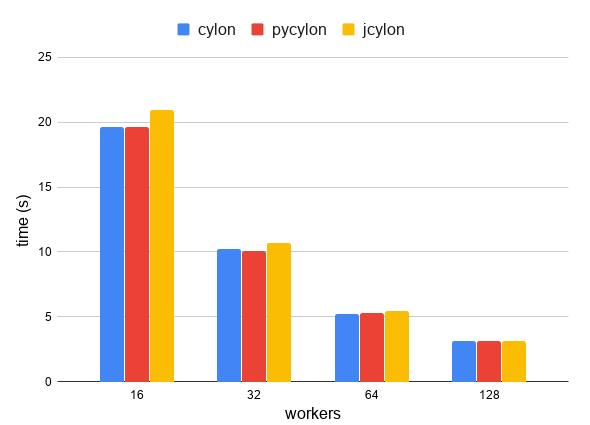}
\end{center}
\caption{Cylon Performance Comparison on C++, Python and Java (Linear on vertical axis and labels on horizontal axis )}
\label{fig:cylon-lang-perf}
\end{figure}

Figure \ref{fig:cylon-lang-perf} shows the time taken for Inner-Join (Sort) for 200 million rows while changing the number of workers. It seems clear that the overheads between Cylon and its Cython Python bindings and JNI Java bindings are negligible. This observation seems to confirm that having a C++ back-end greatly reduces overhead in switching between multiple language runtimes. 

\section{Related Work}\label{s:related-work}

In the data science domain, a widely used data abstraction is the dataframe. Although it has risen to prominence with the advent of deep learning, it was originally developed by S programming language in 1990\cite{s-program-dataframe}. For modern day data science, Python programming language affords rapid prototyping capability. Adopting this, Pandas\cite{mckinney2011pandas} was introduced as a state-of-the-art data representation format. It is a full-fledged Python development based on tabular data. Pandas compute kernels are limited to run only in a single core. With the evolution of big data, an extensive amount of data is being added to data storage every day. For efficient data preprocessing, state-of-the-art big data frameworks like Apache Spark\cite{apache-spark,karau2015learning} also offer a dataframe abstraction on top of the SQL-based Spark DataSet API. PySpark provides integration between JVM-based data structures and Python-based dataframes. Spark also scales in large-scale big data clusters. One of the main challenges in using PySpark is its function as a framework and not as a library. A data scientist needs to set up a Spark cluster separately to run the data pre-processing workloads. The challenge of data movement from JVM to Python also adds a bottleneck in iterative data preprocessing. Modin \cite{petersohn2020towards} provides a Pandas API which can run in large scale with Dask\cite{rocklin2015dask} and Ray\cite{ray-framework} back-ends. Dask includes a distributed dataframe written on Python. Apart from these CPU dataframes, CuDf, a GPU dataframe, was also introduced by Rapids AI. Unlike existing CPU-based solutions, CuDf\cite{rapidsai10:online} utilizes high performance kernels specific for GPUs written in C++. These kernels are exposed to Python via Cython bindings. In addition, frameworks like PyCOMPS\cite{tejedor2017pycompss} along with Dislib\cite{cid2019dislib} provides a better support for distributed computation on array data structures. To optimize an existing Python code, libraries like Numba\cite{lam2015numba} and Pythran\cite{guelton2015pythran} provide high performance capability by optimizing the user code. But in writing frameworks, libraries like Cython\cite{behnel2011cython} and Pybind11\cite{jakob2017pybind11} are recommended to obtain efficient language bindings. By improving the usability for data analytics, Jupyter\cite{kluyver2016jupyter} Notebooks is also widely popular. Finally, in terms of distributed computations, IPyParallel\cite{ipythoni44:online} allows for parallel computation on IPython\cite{perez2007ipython} kernels. IPyParallel is also compatible with Jupyter Notebooks.

\section{Conclusion}\label{s:conclusion}

The exponential growth of data and deep learning applications means it is vital to provide effective data engineering solutions. After studying the existing data engineering solutions and best practices, we showcased how data engineering can be reinforced to get better performance and scale. One of the major qualitative requirements of data engineering is to write ETL pipeline in Python and retain high performance. Using high performance compute kernels written in C++ and offering Cython-based Python APIs means less overhead across the two runtimes and good scaling in HPC environments. We also show how to use data science best practices and extend distributed data engineering towards Jupyter Notebooks. 

From our experiments, we can confirm that Cylon's approach of using Python language bindings with a C++ back-end significantly reduces the overheads of switching between language runtimes. As such, data engineering frameworks could benefit from both Python's convenience and HPC back-end's performance. Furthermore, we showcased that PyCylon's operations scale better than the popular Python-based data engineering frameworks that are available at the time of writing this paper. We firmly believe that PyCylon performance could be improved further by paying careful attention to memory and network utilization and usage of HPC kernels for computations. 

\section{Future Work}\label{s:future-work}

We agree with Petersohn et al's \cite{petersohn2020towards} suggestion that confirming to the Pandas dataframe API is an important feature for Python data engineering tools. We are currently developing a dataframe API based on Modin, and thus Cylon would be another distributed back-end for Modin. To expand our compute kernels we are currently focusing on the supporting distributed computing on array data structures. In supporting diverse data formats, we will be integrating HDF5 and Parquet data loading and data processing in a future software release. Additionally, we don't support GPFS or Lustre file systems yet, but we will be focusing on expanding data storage support in the next stages of the development. Furthermore, we are improving Cylon kernels to be NUMA and cache-aware. We believe this would significantly enhance Cylon's performance. We have also recognized recent developments in communication technologies such as RDMA and Infiniband-enabled message passing without the involvement of the CPU. Here, our focus is to integrate the Cylon communication layer with UCX\cite{ucx}. This permits more flexibility for computation and communication overlap.

\section*{Acknowledgment}
This work is partially supported by the National Science Foundation (NSF) through awards CIF21 DIBBS 1443054, nanoBIO 1720625, CINES 1835598 and Global Pervasive Computational Epidemiology 1918626. We thank Intel for their use of the Juliet and Victor systems, and extend our gratitude to the FutureSystems team for their support with the infrastructure. 

\balance
\bibliographystyle{bibliography/IEEEtran}

\bibliography{bibliography/IEEEexample}

\end{document}